\documentclass[preprint]{aastex}
\def\plotfiddle#1#2#3#4#5#6#7{\centering \leavevmode
\vbox to#2{\rule{0pt}{#2}}
\includegraphics{#1}}
\newcommand{\ltsimeq}{\raisebox{-0.6ex}{$\,\stackrel
        {\raisebox{-.2ex}{$\textstyle <$}}{\sim}\,$}}

\slugcomment{Submitted to AJ 2000 April 11}
\shorttitle{The Light Curve of Aql X-1}
\shortauthors{Welsh et al.}
\begin{document}

\title{The Orbital Light Curve of Aquila X-1}

\author{William F. Welsh, Edward L. Robinson and Patrick 
Young\altaffilmark{1}}
\affil{Department of Astronomy and McDonald Observatory, University of
Texas at Austin, Austin, TX  78712}
\altaffiltext{2}{Department of Astronomy and Steward Observatory, 
University of Arizona, Tucson, AZ 85721--0065} 

\begin{abstract}
We obtained R-- and I--band CCD photometry of the soft X--ray
transient/neutron-star binary Aql~X-1 in 1998 June
while it was at quiescence.
We find that its light curve is dominated by ellipsoidal
variations, although the ellipsoidal variations are severely distorted
and have unequal maxima. 
After we correct for the contaminating flux from a field star 
located only 0\farcs46 away, the peak-to-peak amplitude of the
modulation is $\approx 0.25$ mag in the R band, 
which requires the orbital inclination to be greater than  $36^\circ$. 
The orbital period we measure is consistent with the
18.95~h period measured by Chevalier \& Ilovaisky (1998).
During its outbursts the light curve of Aql~X-1 becomes single humped.
The outburst light curve observed by Garcia et al.\ (1999) 
agrees in phase with our quiescent light curve. 
We show that the single humped variation is caused by a  
``reflection effect,'' that is, by heating of the side of
the secondary star facing towards the neutron star.
\end{abstract}

\keywords{
stars: neutron --- stars: variables --- (stars:)binaries: close
}
\section{Introduction}

Soft X--ray Transients (SXTs, also called X--ray nova) 
are distinguished from other X--ray binaries by their
transient outbursts and soft X--ray spectra (Tanaka \& Lewin 1995;
Tanaka \& Shibazaki 1996; Chen, Schrader \& Livio 1997).
During the outbursts, which recur irregularly at intervals
of 1 to 50 years, their X--ray luminosity increases by a factor of
$10^4$--$10^6$ in a few days and then decays back to minimum on an
e-folding time scale of $\sim 30$ days.
There is a concomitant increase in their optical luminosity
by a factor of $\sim 10^2$.
The leading model for the outbursts is the Disk Instability 
Model (e.g., Cannizzo et al.\ 1998). 
In this model, gas transferred from the normal star in the
binary builds up in a disk around the the compact star.
As the temperature and optical depth in the disk increase, the
viscosity increases, eventually setting off a global instability
that increases the mass flow through the disk,
releasing accretion energy and causing the X--ray outburst.
Between the X--ray outbursts, when the X--ray flux is low,
the masses and dimensions of SXTs can be measured with some confidence.
A startling result of these measurements has been the discovery
that the compact star is a black hole in $\sim$75\% of the SXTs 
(e.g. see Charles 1998).

Aql~X-1 is a particularly interesting SXT because of its
frequent outbursts --- the recurrence time is 
$\sim$ 1 year (Priedhorsky \& Terrell 1984; Kitamoto et al. 1993) -- and
because its X--ray light curve displays richly complex
behavior:
(1) It shows ``type I'' X--ray bursts, demonstrating that the
compact star is a neutron star
(Koyama et al.\ 1981, Zhang et al.\ 1998b, Yu et al.\ 1999);
(2) Nearly coherent oscillations at $\sim$548 Hz were detected during
one type I burst and are likely related to the spin period of the
neutron star (Zhang et al.\ 1998b);
(3) kHz QPOs have been discovered with frequencies in the range 670--870
Hz  (Zhang et al.\ 1998b; Yu et al.\ 1999);
(4) A hard power--law tail extending to at least $>$100 keV has been
detected during outbursts (Harmon et al.\ 1996); and
(5) A sudden spectral hardening of the X--rays has been observed
during the decline from an outburst and has been interpreted as the onset
of a ``magnetic propeller'' state (Zhang et al.\ 1998a; Campana 
et al.\ 1998). 
The inferred strength of the magnetic field,
$\sim1$--$6 \times 10^{8}$~Gauss, is remarkably low for a neutron star. 
The spectral hardening may also be enhanced by the transition into an ADAF
state (Shahbaz et al.\ 1998b; Zhang et al.\ 1998a).

The optical counterpart to Aql~X-1 is V1333~Aql 
(Thorstensen, Charles \& Bowyer 1978).
Two estimates for its orbital period have appeared in the
literature, 18.9479 h (Chevalier \& Ilovaisky 1991, 1998)
 and 19.30 h (Shahbaz et al.\ 1998a). 
The former was derived from the outburst light curve, which has
a single maximum per orbit;
the later period was derived from the quiescent light curve, 
which has two humps per orbit. 
The single hump seen during outburst has been interpreted  as
being due to reprocessed X--rays in an asymmetric disk and/or
the secondary star. 
The double hump modulation seen in quiescence is
interpreted as ellipsoidal variations from the Roche--lobe filling
secondary star. 
If correct, the difference in periods is puzzling;
photometric modulations caused by a non--axisymmetric disk during
an outburst tend to have a longer period than the orbital period. 

Garcia et al.\ (1999) obtained V--band photometry of Aql~X-1 in outburst
and K$^{\prime}$ photometry in quiescence. 
Their outburst data confirm the strong ($\sim$0.6 mag) 
single--humped light curve seen by Chevalier
\& Ilovaisky (1991) and Robinson \& Young (1998), though their data were
not sufficient to refine the orbital period. 
The quiescent K$^{\prime}$ photometry did not show any clear evidence 
for a modulation, suggesting a low orbital inclination. 
They also obtained spectroscopy in quiescence that
revealed a weak, low--amplitude modulation of
the absorption line radial velocities, 
also implying a low orbital inclination. 
While this agrees with the analysis of the quiescent photometric 
modulations by Shahbaz et al.\ (1998a), it is again puzzling: 
the large amplitude
modulations seen in the outburst state suggest a high inclination.
Furthermore, Garcia et al.\ (1999) note that the relative phasing of the
outburst maxima they observed  and the quiescent minima from the Shahbaz
et al.\ study disagree. 
Clearly, the interpretation of the modulations in
the light curves is in a confused state.

Some of this confusion was resolved when 
Callanan et al.\ (1999) obtained high-resolution images of
Aql~X-1 showing that the object previously thought to be Aql~X-1
is actually two stars separated by only 0\farcs46.
Chevalier et al.\ (1999) subsequently showed
that the {\em fainter} of the two stars is the counterpart of Aql~X-1,
and that 88\% of the V--band light is from the line--of--sight
contaminating star. 
The presence of a large amount of contaminating light
has two immediate consequences: 
(1) The fractional amplitude of any
photometric variations have been substantially underestimated; this leads
to an underestimate of the inclination based on the amplitude of the
ellipsoidal variations; and 
(2) previous determinations of the
spectral type of the secondary star (K0 by Thorstensen et al 1978; K1 by
Shahbaz et al. 1997) are incorrect. 
Based on its photometric colors, the spectral type of the secondary
star is K6--M0 (Chevalier et al.\ 1999).

In this paper we present new optical photometry of Aql~X-1 obtained
during quiescence, and use these data to revise the orbital period
and constrain the inclination of the binary system. In \S2 we describe
the data acquisition and calibration. In \S3 we derive a revised orbital
period from the light curve modulations, and compare it with previous
published values.
Using the estimates of Chevalier et al.\ (1999), we then correct the light
curves for the contaminating light from the nearby field star.
In \S4, the contamination--corrected quiescent light curves are 
modeled as ellipsoidal variations and an estimate for the binary system
inclination is derived. The outburst light curve of Garcia et al.\ (1999)
is also investigated; we model the outburst photometry with a
reflection/heating effect of the companion star.
Section 5 summarizes our results.

\section{Observations and Data Calibration}

We observed Aql~X-1 on 1998 June 24 -- 28 UTC.
Figure~1 shows the X--ray light curve of Aql~X-1 in 1998, obtained 
with the All Sky Monitor on the {\em Rossi X--Ray Timing Explorer}
\footnote{ASM light curves are publically available at
http://space.mit.edu/XTE/ASM\_lc.html and at
http://heasarc.gsfc.nasa.gov/docs/xte/asm\_products.html.},
with the times of our observations marked by arrows.
Aql~X-1 erupted in 1998 March but our observations began
more than 40 days after the end of the eruption
and Aql~X-1 was fully in quiescence.

We measured the optical light curve at the Cassegrain focus of
the 2.1-m telescope at McDonald Observatory using a Tektronix CCD, which
produced an image scale of 0\farcs352/pixel. The observing pattern was
10 images in the R band followed by 3 images in the I band,
repeated \textit{ad paucas noctis}.
All exposures were two minutes in duration, yielding an integrated
S/N of $\sim100$ in  the R band and $\sim 75$ in the I band.
Observing conditions were generally good, with dark skies and
1\farcs5 -- 2\farcs0 seeing, although occasional patchy clouds and smoke
from nearby forest fires were present.
The telescope tracking was, however, poor, and contributed
significantly to the image point--spread function.
Frames with poor image quality ($>$2\farcs08) proved to be unusable
because of contamination from nearby field stars, 
resulting in the loss of roughly a third of the data.
We obtained a total of 312 usable images in the R band and 85 
in the I band.

We extracted instrumental magnitudes for Aql~X-1 and four local 
comparison stars using the DAOPHOT routine in IRAF (Stetson 1987, 1990).
The local comparison stars did not vary in our data, nor are they
listed as variable stars by Reynolds, Thorstensen \& Sherman (1999)
in their study of 6104 stars near Aql~X-1.
Aql~X-1 lies in a very crowded field and is separated by only
2\farcs2 and 2\farcs4 from two nearby field stars (see Figure~2);
and since these stars are only two and three magnitudes fainter than
Aql~X-1, it was essential to extract the flux from Aql~X-1 by fitting
a point spread function to the images. We used two iterations of the
DAOPHOT point spread function fitting routine; 
a third iteration did not improve the precision of the photometry.
The field star separated from Aql~X-1 by 
only 0\farcs46 is not resolved from Aql~X-1 in our data.  
Our measurements give the sum of the flux from this star 
and Aql~X-1 (called stars \textit{e} and \textit{a} 
by Chevalier et al.\ [1999]).
We calibrated the fluxes by observing Landolt standards
(Landolt 1992) and then used the conversions from R and I magnitudes 
to fluxes given in Lamla (1982).
The RMS scatter in the final measured fluxes from the local 
comparison stars is $< 0.56$\%.

We emphasize that the use of point spread functions to extract
the flux from Aql~X-1 was crucial to the success of this project.
The amplitude of the variations of the combined flux from Aql~X-1 
and its 0\farcs46 companion is small, and ordinary 
aperture photometry would have allowed significant and variable 
contamination from the other nearby stars, producing unacceptably
large errors in the flux measurements.

\section{The Light Curves}
\subsection{The Orbital Period}
The observed R--band light curve varies by about 
0.05 mag peak--to--peak. 
We searched for periodicities in the light
curve using an assortment of methods, including the phase--dispersion
minimization method, discrete Fourier transforms, Lomb--Scargle
periodograms, and ``clean'' power density spectra.\footnote{
The PDM, clean and Lomb--Scargle periodogram algorithms were contained
in the software ``PERIOD'' V4.2 written by V.S. Dhillon and provided
by Starlink: http://star-www.rl.ac.uk/cgi-store/storetop.}
All showed a multiplicity of periods, the strongest being near 0.6,
1.6, 2.3 and 2.6 cycles per day. 
However, it was obvious that the
effects of a 1--day sampling alias dominate the analysis.
Figure 3 shows the low--frequency portion of the power spectrum calculated
from the R--band light curve. 
Nearly all features in the power spectrum can be ascribed to aliases of
the fundamental frequency and its harmonics. 
The ``tree'' diagram in the lower panel of the figure
marks the positions of the fundamental, the first harmonic, and
their 1--day aliases and demonstrates that we account for every 
large peak in the power spectrum with just one 
number: the fundamental frequency.

We used a physically motivated method 
to extract the period of the fundamental.
Assuming that the light curve is dominated by ellipsoidal
variations that produce a signal at the orbital period and half the
orbital period, we fit the light curve with a model consisting of the
sum of two sinusoids whose amplitudes and phases are free, but whose
frequencies are locked at $f_{0}$ and $f_{1}=2f_{0}$, where $f_0$ is the
frequency of the fundamental and $f_1$ is the frequency of the 
first harmonic.
The minimum $\chi^{2}$ of the fit gives the orbital period.
This 2--sine fit produced a periodogram similar to the classical (1--sine)
periodogram and the Fourier transform power spectrum in that it had minima
at nearly the same frequencies, but it contained additional minima and the
minima were substantially different in relative depth, e.g., the lowest
minima in the classical periodogram was near 0.6 c/d, while in the
2--sine fit the minimum was near 1.3 c/d. 
To test how well our model reproduces the observations, 
we subtracted the best fit model from the light curve and
recomputed the power spectrum.
The power spectrum of this
``pre--whitened'' data is also shown, to scale, in Fig 3. 
The remarkable
reduction of power verifies that nearly all the power is contained in
just the fundamental and its first harmonic. 

The R--band light curve yielded a fundamental frequency of 
1.283 cycles/day, 
corresponding to an orbital period of 18.71 $\pm$ 0.06 h.
The upper light curve in Figure~4 shows the R--band light curve
folded at this period along with the best fitting 2--sine fit.
The (marginally) deeper of the two
minima in the folded light curve occured at HJD 2450988.8204, which we
define as phase zero. 
All phases in this paper use this period and 
definition of phase zero.
The reduced $\chi^{2}$ of the fit was 1.27 with 306 degrees of freedom.
The $1\sigma$ uncertainties were derived from the period interval
encompassing the reduced  $\chi^{2}_{min}$ + 1.
To check the robustness of this period measurement,
we recomputed the fit giving all data points equal weight, and then again
after removing three apparent outliers:
The period changed by less than 0.02 h.
As a further test, we fit the comparison star light curve 
with the 2--sine model. 
When forced to have a frequency $f_{0}$ near 1.283 c/s (the frequency
found in the Aql~X-1 light curve), the individual amplitudes of the 2
sines in the model fit were $<$0.4\% and $<$0.2\%; in contrast, the
amplitudes for the 2 sines when fit to the  Aql~X-1 light curve were
both $\sim$1.3\%.

The I--band light curve has fewer and noisier data points, and
the $\chi^2$ of the 2--sine fit was worse $(\chi^{2} = 2.77)$.
Nevertheless, the period derived from the I--band light curve,
18.85$\pm$ 0.06 h, is not much different from the
R--band estimate, and the formal uncertainty is the same
because the amplitude of the modulation is larger in the 
I--band than in the R--band.

\subsection{Comparison with Previous Determinations of the Orbital Period}

The orbital period determined from our data, 18.71$\pm$0.06 h, 
is $\sim$1.3\% shorter than the 18.9479~h period found
by Chevalier \& Ilovaisky (1998).
While this is nearly a 4$\sigma$ difference, we do not consider the
difference significant because the quoted uncertainty in our period
does not include systematic errors introduced by, for example, 
intrinsic variability, a non--white power spectrum, or the one-day
aliasing.

Chevalier \& Ilovaisky (1998) also found that additional photometry taken
while \mbox{Aql~X-1} was in quiescence agrees with the orbital period
determined from outburst data to within 0.02\%. 
They conclude that the
period determined using outburst data is the orbital period. 
Our result supports this interpretation.

If we use the time of the minimum of the outburst light curve given by
Garcia et al. (1999) as the definition of phase zero (inferior
conjunction of the secondary star) and the orbital period given by
Chevalier \& Ilovaisky (1998), we can project their ephemeris onto our
observations and compare the relative phasing. Using $T_{0}$=2450282.220
$\pm$0.003 HJD and period 0.789498$\pm$0.000~010 d, we find that 
the observed time of minimum in our data, HJD 2450988.8204, is at
phase 0.000$\pm$0.012, in excellent agreement with expectations.
This agreement in phasing further supports our conclusion that the
orbital period is 18.9479 hours.

The photometric period reported by Shahbaz et al.\ (1998a), 
19.30$\pm$0.05 h, cannot be the orbital period of Aql~X-1.
Is it a superhump period?
Superhumps are quasi--periodic modulations in the outburst light curves, 
most likely caused by a precessing elliptical accretion disk and excited
by a tidal resonance with the secondary star (see Warner 1995). 
Superhump periods are typically a few percent
larger than the orbital period in the SU~UMa cataclysmic variable stars.
If one naively extrapolates the orbital
period--superhump period relation for the SU UMa stars
to Aql~X-1, the predicted difference between the orbital period
and superhump period is $\sim$50\%, far greater than the
observed 1.9\% difference. 
Furthermore, the superhump models require the
accretion disk to extend out to the 3:1 tidal resonance, and this is only
possible for binaries with extreme mass ratios, $q \ltsimeq 0.22$. 
For Aql~X-1, we can estimate a value for $q$ by assuming the primary is
a neutron star with mass 1.4$M_{\sun}$ and the secondary is a K7 V star
(Chevalier et al.~1999) with mass 0.6$M_{\sun}$. 
This gives $q=0.43$, far above the upper limit allowed for the 
tidal resonance that excites superhumps. 
Even for a mass ratio near $q$=1/3,
which we prefer because this is more appropriate for an evolved secondary
star that fills its Roche lobe, the mass ratio remains too high for the
superhump interpretation. 
Thus it is unlikely that the period difference can be ascribed to
a superhump phenomenon.
Given the small amplitude of the variations in the light curve
of Aql~X-1, the crowded field, and the strong 1-day aliases, we suspect
that the 19.3~h period is spurious.

\subsection{The Uncontaminated Light Curve}

The folded light curve at the top of Figure~4 includes flux from
the field star 0\farcs46 away from Aql~X-1.
According to Chevalier et al.~(1999)
the field star contributes 88\% of the flux in the V band and 
77.5\% of the flux in the I band.
To correct the I--band light curve for this contamination
we simply subtracted 77.5\% of the mean flux from the light curve.
To correct the R--band light curve we linearly interpolated 
between the V-- and I--band magnitudes of the field star given 
by Chevalier et al.~(1999) and then applied a correction of
0.105 $\pm$0.015 mag to compensate for the intrinsic color of the
field star (estimated to be a late--G to early K type star
by Chevalier et al.~1999). 
We estimate that 83$\pm 2$\% of the flux in the R--band is
contamination from the field star, where the estimated
standard deviation comes partly from the estimated measurement 
error and partly from uncertainty about the orbital phase and 
therefore the brightness of Aql~X-1 at the time of the measurements.
The lower light curve in Figure~4 is the R--band light curve after
the flux from the field star has been removed. 
The amplitude of the modulation in the uncorrected
R--band light curve is $\sim$4.2\% peak--to--peak,
but after subtracting the contribution from the contaminating star, the
modulation amplitude jumps to $25\pm3$\%. 

\section{Modeling the Orbital Modulation}
\subsection{The Quiescent Light Curve}

The quiescent light curve shown in Fig.~4 has two humps per orbital
period, which we interpret as ellipsoidal variations of the late K star. 
The ellipsoidal variations are, however,
clearly distorted since the two humps have unequal maxima.
The source of the distortion is unknown but there are
two likely candidates, star spots on the K star and non-axisym\-metries
in the accretion disk around the neutron star.
With few exceptions the dominant effect of both distortions is to 
increase the amplitude of one ellipsoidal hump with respect to the other. 
Thus, fitting an ellipsoidal light curve to the larger hump 
is likely to
overestimate the true amplitude of the ellipsoidal variation and
will, therefore, yield an upper limit to the orbital inclination.
Conversely,
fitting an ellipsoidal light curve to the smaller hump will 
yield a lower limit to the inclination.

We fit the contamination--corrected R--band light curve of Aql~X-1 with
synthetic light curves calculated from an updated and rewritten version
of the light curve synthesis program discussed by Zhang et al.\ (1986). 
We adopted a blackbody spectral distribution with a temperature of 4000~K 
for the late K star, and
used a gravity darkening coefficient of \( \beta = 0.08 \) and linear
limb darkening coefficients of \( u = 0.8 \) 
and \( u = 0.7 \) for the R--band and I--band light curves respectively
(Claret 1998). The derived orbital inclinations are relatively insensitive
to any of these parameters. 
Changing the temperature to 5000~K, for example,
changed the derived orbital inclinations by only 2 -- $3^\circ$.
The ratio of the masses, $q = M_K / M_{\rm n}$, 
must be less than 1 since we do not observe dynamically unstable
mass transfer, and it must be greater than 0.1 because the
mass accreting star is a neutron star.
If the K star is a stripped giant, the mass ratio should be
close to $q = 1/3$. 
The abundant evidence for mass transfer demands that the K star
fill or nearly fill its Roche Lobe.

Figures~5 and 6 show the fits of synthetic light curves to the R--
and I--band light curves for models in which $q = 1/3$ and the K star
is the sole contributor to the light curve.
The orbital inclinations derived from the R--band light curve
are $36^\circ$ and $55^\circ$ for the smaller and larger humps
respectively, and the inclinations from the I--band light curve are
$58^\circ$ and $75^\circ$.

For several reasons we believe that the orbital inclination derived 
from the fit 
to the smaller hump in the R--band light curve is the most robust 
of the inclination measurements, but that it should be interpreted 
only as a lower limit to the orbital inclination. 
First, the quality and the quantity of the R--band data are much higher 
than the I--band data, so we strongly prefer the
fits to the R--band light curves.
Second, although the inclinations derived from fits to the larger
hump depend fairly sensitively on the adopted mass ratio, the
inclinations derived from the small hump are relatively insensitive.
The inclination derived from the small hump in the 
R--band light curve ranged only from $33^\circ$ to $41^\circ$ over 
the extreme range of permissible mass ratios.
Third, the accretion disk certainly contributes some of the
flux from Aql~X-1 in the R band, if for no other reason than
H$\alpha$ emission is observed in the spectrum of Aql~X-1.
(Note that the true equivalent width of the H$\alpha$ emission line
becomes almost 6 times greater than the observed equivalent width
when flux from the contaminating field star is removed; the equivalent
width is boosted from $\sim$5.2\AA \ [Shahbaz et al.\ 1997; 1998b] to
$\sim$30\AA). The effect of ignoring the disk emission is to make the
derived orbital inclination too small.

We conclude that the orbital inclination of Aql~X-1 is greater than
$36^\circ$. This inclination is greater than the upper limits of
$31^\circ$ given by Garcia et al.\ (1999) and 20$^\circ$--30$^\circ$
(90\% confidence interval) given by Shahbaz et al.\ (1998a). 
The limits derived in these papers were based on modeling the 
ellipsoidal variations in the quiescent K$^\prime$--band and I--band 
light curves, respectively. 
Setting aside the fact that the variations were only marginally detected
in the light curves, there is a simple explanation for the
discrepancy:  We have removed the contaminating flux from the field star
0\farcs46 away from Aql~X-1.
When the contaminating flux is removed, the amplitude of the
ellipsoidal variations increases substantially, and with it the derived
inclination.

Garcia et al.\ (1999) derived an upper limit of
$12^\circ$ for the orbital inclination from their measurement of
the upper limit to the amplitude of any radial velocity
variations of the secondary star ($<$48 km s$^{-1}$) and the assumption that
Aql~X-1 is similar to Cen~X-4, a neutron star SXT. 
This upper limit is much lower than our lower limit on the inclination. 
We believe this discrepancy is again due to the contaminating flux
from the field star:  The (presumably) non--varying radial velocity 
of the contaminating star biases measurements of orbital radial 
velocity variations towards lower values.
Likewise, because the contaminating star is probably a late G--type
star (Chevalier et al.\ 1999), the lines from the contaminating star and
the secondary star in Aql~X-1 will blend, thus making any measurement
of the rotational broadening of the absorption lines highly suspect. 
Shahbaz et al.\ (1997) found a rotational broadening of the secondary star's
absorption lines of $V_{rot} \sin \ i = 62^{+30}_{-20}$ km s$^{-1}$.
Since the resolution of their spectrograms was 165 km s$^{-1}$ and
since the spectrum is bended, this value must be treated with skepticism.
Also, the non--detection of the Li 6707 \AA \ line by Garcia et
al.\ (1999) is not significant --- if one boosts their detection threshold
of $\sim$0.1 \AA \ equivalent width by a factor of $\sim$6 to compensate
for the contaminating flux, then the equivalent width 
of lithium in Aql~X-1 is consistent with that found for Cen~X-4.

Shahbaz et al.\ (1997) found that the full--width at half--maximum of the
H$\alpha$ emission line is broader in \mbox{Aql~X-1} than in Cen~X--4.
The FWHM should not be significantly biased by the presence of the
contaminating star, so inclinations based on velocity widths 
should remain valid.
They derived an inclination of $\sim$50$^\circ$, which is entirely
consistent with our measurement. 
As noted by Shahbaz et al.\ however, any gas with non--Keplerian 
motions that contributes to the emission line in Aql~X-1 would
weaken this argument.
Finally, Garcia et al.\ (1999) estimated that the inclination must be 
roughly $\sim$70$^\circ$ to account for the large amplitude of the
variations in the outburst light curve of Aql~X-1.
We conclude that the orbital inclination of Aql~X-1 cannot be small
and, indeed, must be greater than $36^\circ$.

\subsection{The Outburst Light Curve}

Figure~7 shows the V-band light curve of Aql~X-1 obtained 
by Garcia et al.\ (1999) in 1996 July during a weak, flat-topped outburst.
In contrast to the light curve at minimum light, the outburst light
curve has a single maximum (for additional examples, see Robinson \&
Young 1997 and Chevalier \& Ilovaisky 1991). 
As noted earlier, if we adopt the orbital period
of Chevalier et al.\ (1998), we find that the minimum of the V-band
light curve is precisely in phase ($\Delta\phi < 0.012$) with the 
$\phi = 0$ minimum of our quiescent  light curve, which was obtained two
years later. This strongly suggests that the outburst light curve is
caused by heating of the secondary star by flux from the neutron star and
accretion disk.

Because the outburst light curve was obtained over several nights and
because it is distorted by flickering, fits of refined models to the
light curve are unwarranted.
Instead we have calculated a synthetic light curve from a
a simple model in which a late K--type star fills its Roche lobe and is 
heated by flux from the neutron star.
The model has an orbital inclination of $40^\circ$, a mass ratio 
of $q = 1/3$, and a luminosity ratio $L_{\rm n}/L_K \approx 20$.
We have no information about the scale of the binary system, but 
if the luminosity of the highly irradiated K star is $10\ L_\odot$, 
the luminosity of the neutron star would be 
$L_{\rm n} \approx 10^{36}\ \textrm{ergs s}^{-1}$, not an unreasonable
luminosity for this outburst.

The fit of the model to the light curve is shown in Figure~7.
While the fit is hardly perfect, the model successfully
reproduces the shape, amplitude and phasing of the observed
light curve. Furthermore, the inclination is consistent with the
inclination derived from the ellipsoidal variations, thus solving the
puzzle noticed by Garcia et al.\ (1999): the large amplitude of the
variations seen during outburst requires a much higher orbital
inclination than was originally inferred from fits to the ellipsoidal
variations in the quiescent photometry (though an inclination as large
as $\sim$70$^\circ$ is not necessary).
We therefore conclude that the optical light curve of Aql~X-1 was
dominated by heating of the secondary star during the July 1996 eruption.

\section{Summary}

We have confirmed the presence of a double--humped orbital modulation 
in the quiescent light curve of Aql~X-1. 
Our orbital period estimate of 18.71~h is slightly shorter than,
but consistent with, the 18.95~h period found by
Chevalier \& Ilovaisky (1998).
A single orbital period, 0.789498~d, fits all the available
light curves during both quiescence and outburst.

After subtracting the contaminating flux from a field star
0\farcs46 away from Aql~X-1, the peak-to-peak amplitude of the
double--humped modulation is $\sim 0.25$ mag.
The modulation is caused by ellipsoidal variations of the
K star in the system.
The ellipsoidal variations are distorted but do, nevertheless,
yield a lower limit to the orbital inclination of Aql~X-1.
We find that the inclination must be greater than $36^\circ$.

In outburst, the orbital light curve contains a single, large--amplitude 
hump. A synthetic light curve produced by an irradiated, late--K star 
provides a good match to the outburst light curve observed by Garcia
et al. (1999). Further support for the heated secondary star model comes
from the excellent phase agreement between the outburst and quiescence
photometric variations. During outbursts, the optical light is most
likely dominated by the heating of the K--type secondary star.

\acknowledgments

We thank Drs. Mike Garcia and Paul Callanan for their outburst photometry
and for general helpful discussions.
This work was supported in part by the NSF through grant AST--9731416.
This paper made use of results provided by the {\it ASM/RXTE} teams at MIT
and at the {\it RXTE} SOF and GOF at NASA's GSFC.


\newpage

\clearpage
\begin{figure}
\plotfiddle{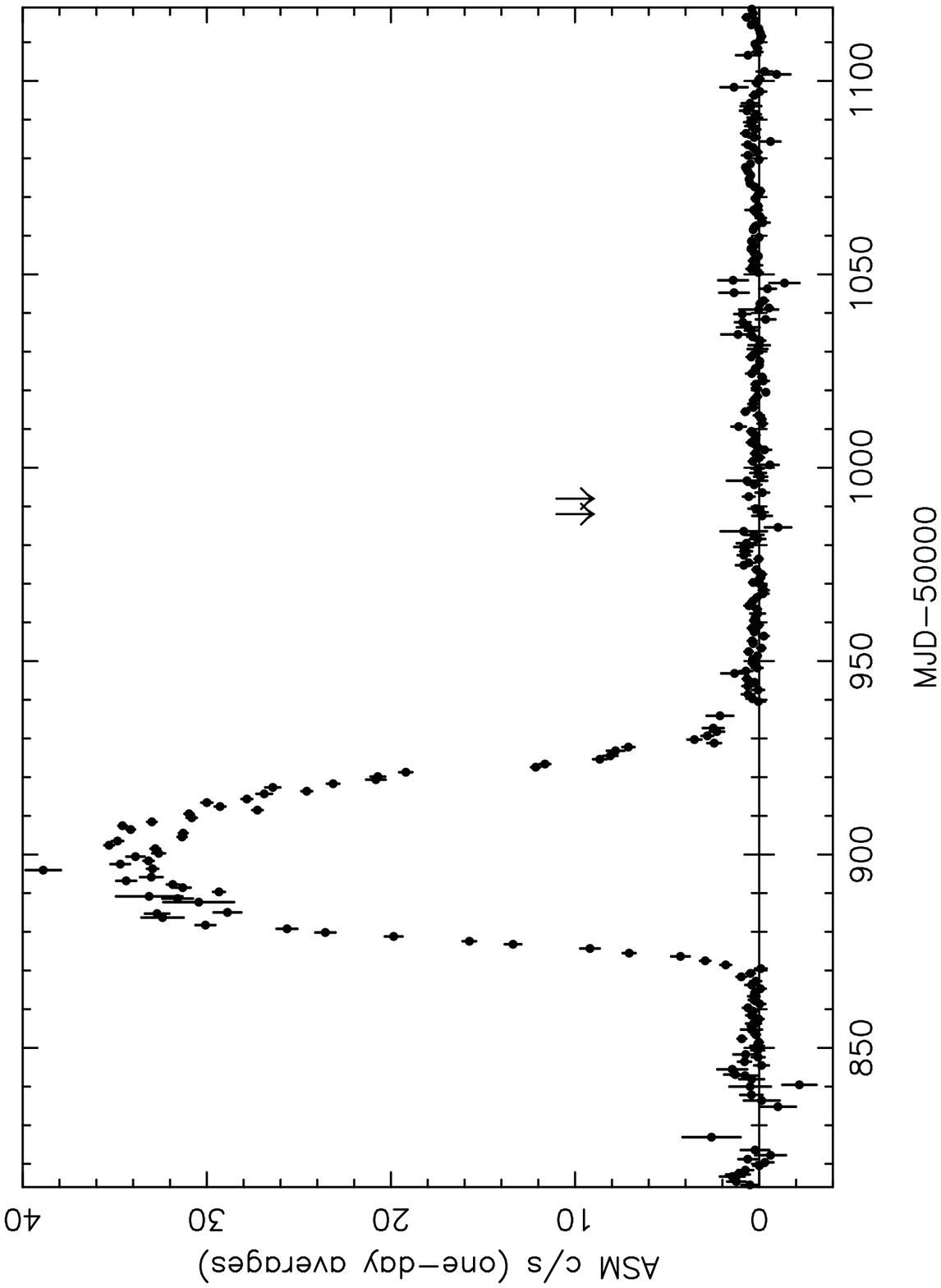}{3.0in}{-90}{50}{50}{-200}{350}
\caption{ 
The {\it RXTE} ASM light curve of Aql~X-1 showing the X--ray outburst
of 1998 March. Indicated with two arrows are the start and stop times
of our optical photometry observing run.
\label{fig1}}
\end{figure}

\clearpage
\begin{figure}
\plotfiddle{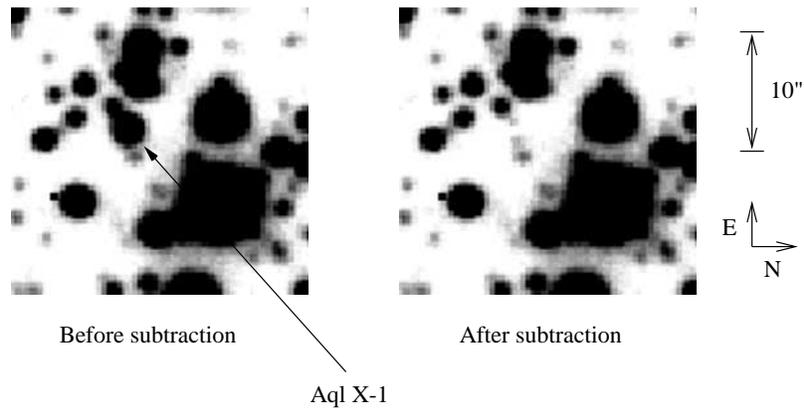}{3.0in}{-90}{50}{50}{-200}{350}
\figcaption{ 
The field of Aquila X-1.  The image on the left
is the sum of 10 R--band CCD frames, each 2 minutes long.  The image on
the right is the same but Aql~X-1 has been subtracted from the image
(via DAOPHOT PSF fitting). The two faint stars $\approx
2^{\prime\prime}$ distant on either side of
Aql~X-1 vitiate simple aperture photometry of Aql~X-1 when it is near 
quiescence.
\label{fig2}}
\end{figure}

\clearpage
\begin{figure}
\plotfiddle{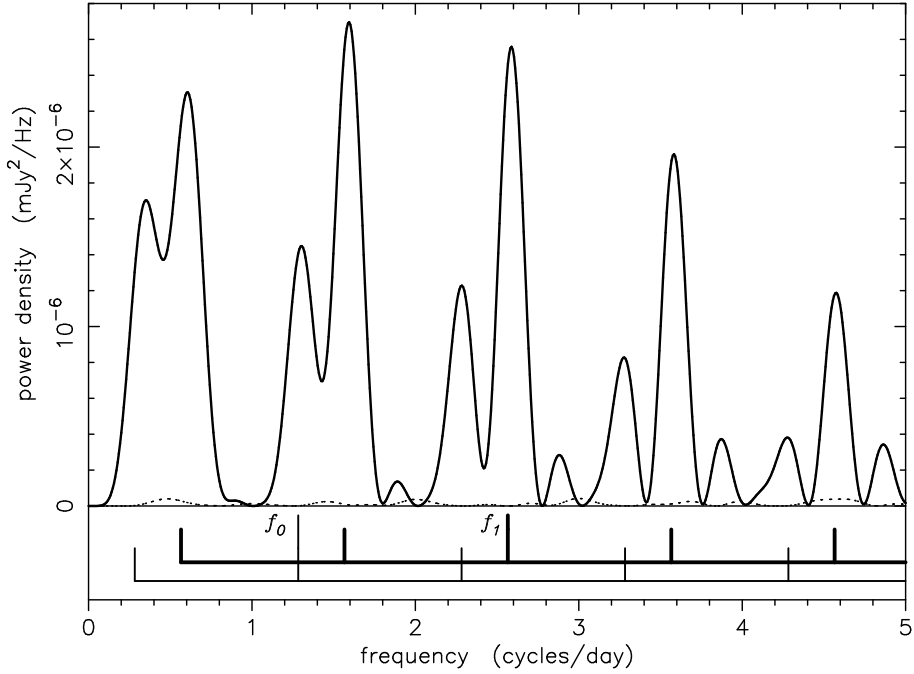}{3.0in}{-90}{50}{50}{-200}{350}
\caption{ 
The power spectrum of the R--band light curve.
The power spectrum has many peaks but the peaks can all be derived from
(1) a fundamental frequency $f_0$, (2) its first harmonic $f_1$, and 
(3) their 24-hour aliases.  Frequency $f_0$ corresponds to the orbital
period  ($\textrm{P} = 1/f_{0}$) and $f_1$ is produced by the
double--humped ellipsoidal variation in the light curve.
When the light curve is prewhitened by removing sine curves with
frequencies $f_0$ and $f_1$, all significant power vanishes. The dotted
line shows the power spectrum of the prewhitened light curve. 
\label{fig3}}
\end{figure}

\clearpage
\begin{figure}
\plotfiddle{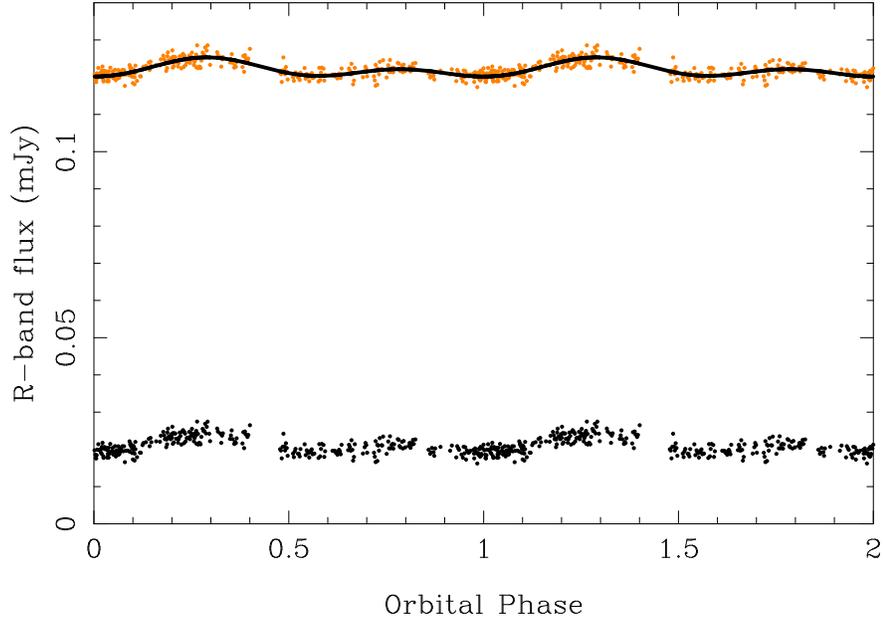}{3.0in}{-90}{50}{50}{-200}{350}
\caption{ 
The R--band light curve of Aql~X-1 folded at the orbital period.
The upper set of points are the observed fluxes including the
contaminating light from the field star $0.46^{\prime\prime}$ away.
The lower set of points is the light curve after the contaminating flux
has been removed. The amount of contamination, 83\% of the total R--band
flux, is very significant and dramatically changes the fractional
amplitude of the orbital modulation. The solid line is a
two--sine--curve fit to the data as explained in the text.
\label{fig4}}
\end{figure}

\clearpage
\begin{figure}
\plotfiddle{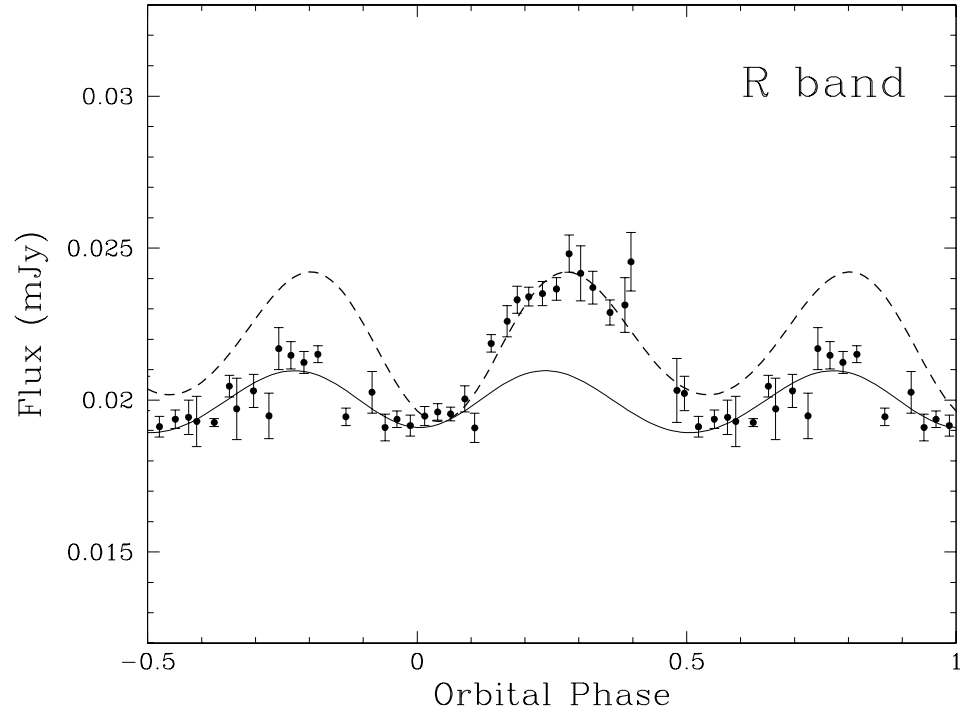}{3.0in}{-90}{50}{50}{-200}{350}
\caption{ 
The orbital phase folded and binned R--band light curve fit with an
ellipsoidal variations model. 
The dashed curve has been fit to phases 0.0 -- 0.5 (the larger hump) and
the solid curve has been fit to phases 0.5 -- 1.0  (the smaller hump).
The two fits require different inclinations  ($36^\circ$ and $55^\circ$
degrees), which bracket the true inclination.
\label{fig5}}
\end{figure}

\clearpage
\begin{figure}
\plotfiddle{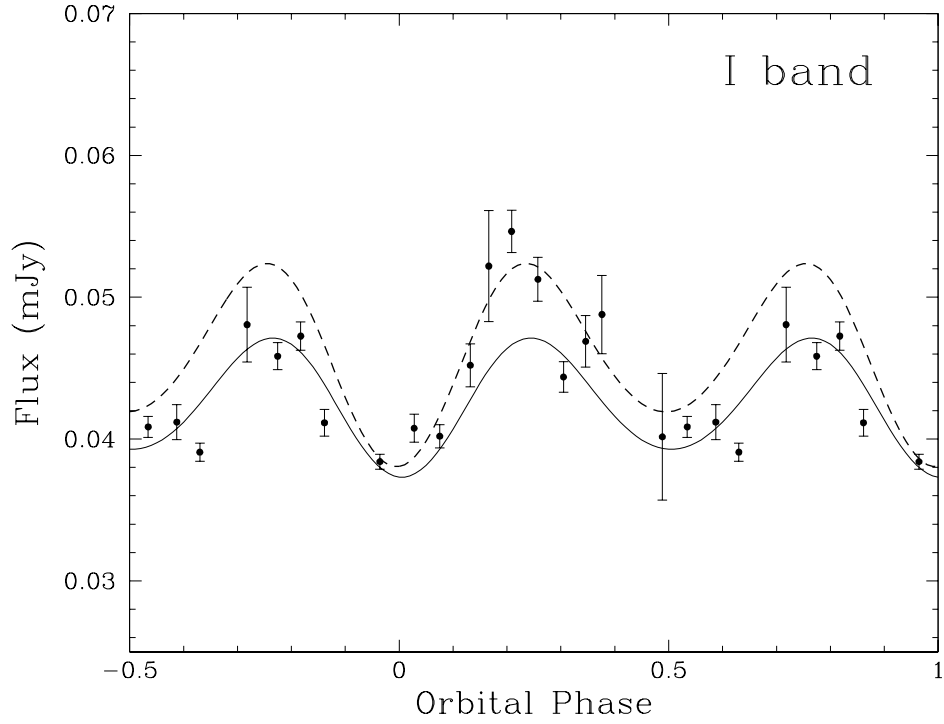}{3.0in}{-90}{50}{50}{-200}{350}
\caption{ 
The orbital phase folded and binned I--band light curve fit with an
ellipsoidal variations model.
The dashed curve has been fit to phases 0.0 -- 0.5 and the solid curve
has been fit to phases 0.5 -- 1.0. The two fits have inclinations of
$58^\circ$ and $75^\circ$, respectively. Compare with the R--band light
curve shown in Fig 5.
\label{fig6}} 
\end{figure}

\clearpage
\begin{figure}
\plotfiddle{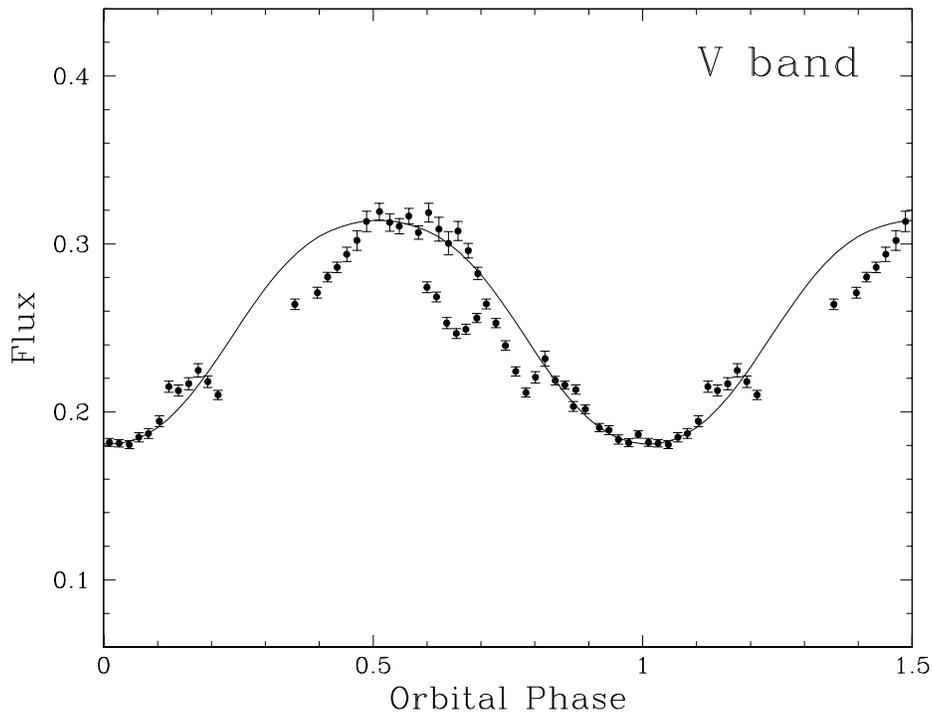}{3.0in}{-90}{50}{50}{-200}{350}
\caption{ 
The V--band light curve of Aql~X-1 during outburst as measured by Garcia
et al.\ (1999). The solid line is the synthetic light curve produced by
an irradiated, late--K star in a binary with an orbital inclination of
$40^\circ$  and a mass ratio of $q = 1/3$.
This representative model light curve is not a formal fit to the
observations. While the match to the data is not perfect, the
irradiated star model can account for the shape, phase, and amplitude of
the light curve.
\label{fig7}}
\end{figure}

\end{document}